# Temperature-Modulated Differential Scanning Calorimetry Analysis of High-Temperature Silicate Glasses

Tobias K. Bechgaard[1,*], Ozgur Gulbiten[2], John C.Mauro[3], Yushu Hu[4], Mathieu Bauchy[4], Morten M. Smedskjaer[1]

[1] *Department of Chemistry and Bioscience, Aalborg University, Aalborg, Denmark*

[2] *Science and Technology Division, Corning Incorporated, Corning, USA*

[3] *Department of Materials Science and Engineering, The Pennsylvania State University, University Park, USA*

[4] *Physics of AmoRphous and Inorganic Solids Laboratory (PARISlab), University of California, Los Angeles, USA*

* e-mail: tkb@bio.aau.dk

## Introduction

Differential scanning calorimetry (DSC) is one of the most versatile probes for silicate glasses, allowing determination of, e.g., transition temperatures (glass, crystallization, melting) and the temperature dependence of heat capacity [1]. However, complications arise for glasses featuring overlapping transitions and low sensitivity, e.g., arising from $SiO_2$-rich compositions with small change in heat capacity during glass transition or the low sensitivity of thermocouples at high temperature. These challenges might be overcome using temperature-modulated DSC (TM-DSC), which enables separation of overlapping signals and improved sensitivity at the expense of increased measurement duration [8]. In TM-DSC a sinusoidal heating curve is superimposed on the linear heating rate known from standard DSC and the temperature ($T$) profile is thus given by:

$$T = T_0 + \beta t + A\sin(\omega t) \qquad (1)$$

where $T_0$ is the initial temperature at time $t = 0$, $\beta$ is the heating rate, $A$ is the amplitude of the modulation and $\omega$ is the angular frequency of the modulation ($\omega = 2\pi/P$, where $P$ is period).

In the glass science community, TM-DSC has been used to analyze organic, chalcogenide, metallic, and certain oxide glasses. Common to all the present studies is the use of glasses with relatively low glass transition temperature ($T_g$ below 600°C), as this has been the temperature limit of commercially available TM-DSCs, thus excluding the majority of industrially relevant silicate glasses.

## Parametric study of TM-DSC for silicate glasses

Recently it has become possible to perform temperature-modulated experiments on commercial instruments at high-temperature (>700°C). However, the transition from low-temperature to high-temperature instruments is not straightforward, as the experimental parameters used for a typical low-temperature TM-DSC experiment cannot be directly transferred to that at high-temperature. Therefore, in order to obtain good data quality, the experimental parameters must be adjusted to fit the probed glass and instrument. The goal when designing the experimental TM-DSC parameters is to obtain a linear response between input and output data, while achieving a high signal-to-noise ratio (S/N). In order to elucidate the effect of varying underlying heating rate, amplitude, and frequency on linearity and S/N, we have performed a systematic parametric study on glass compositions with different $T_g$ and liquid fragility ($m$) [13].

We find that for typical high-$T_g$ calcium aluminosilicate glasses ($T_g \sim 900$°C and $m$ = 20-50), $\beta$ should typically be 2-3 K/min, but for glasses with $SiO_2$ content above ~80 mol%, $\beta$ could be as high as 5 or 7 K/min. $A$ values mainly affect the S/N. On the utilized Netzsch STA 449F1 Jupiter® equipment, amplitudes of 5 K/min are necessary to achieve adequate S/N. Higher amplitude might be useful in special cases, such as for pure silica glass with low $m$, but too high amplitude results in smearing of the data and loss of resolution.

Furthermore, Lissajous curves can be used to evaluate if the data exhibit a linear response between heating rate and heat flow, which is required for deconvolution of the raw data and important when studying glass relaxation processes as large enthalpy releases may occur [14]. We find that the data for the studied high-$T_g$ glasses exhibit linear responses when low heating rates (2-3 K/min) are used in combination with amplitudes that ensure high S/N (Fig. 1).

## Liquid fragility determination using TM-DSC

Angell defined the liquid fragility index $m$ as the slope of the logarithmic viscosity ($\eta$) vs. scaled inverse temperature ($T_g/T$) curve at $T_g$:

$$m(x) = \left.\frac{d\log_{10}\eta(T,x)}{d\left(T_g(x)/T\right)}\right|_{T=T_g(x)} \qquad (2)$$

As the glass transition is a relaxation phenomenon, it is also possible to estimate $m$ from the temperature dependence of the structural relaxation time. This has been done for molecular glasses by determining the dielectric relaxation time [15], but it is not possible for silicate glasses as it only relates to polar atomic motions [16]. Instead TM-DSC has been proposed as a technique for measuring the structural relaxation time of silicate glass-forming systems, using the thermal relaxation caused by temperature modulation to measure the structural relaxation time [17]. That is, the thermal relaxation occurring during modulation in a TM-DSC experiment can be used to investigate and understand the relaxation occurring in melts at high temperature.

The deconvolution of raw TM-DSC data allows for separation of the calorimetric contributions from the enthalpy relaxation during the glass transition and the heat capacity of the glass itself, identified by the imaginary and real heat capacity, respectively [8]. Using the frequency dependence of the peak temperature ($T_g^{\omega}$) in the imaginary heat capacity, TM-DSC has already been used to determine fragility of several glass-forming borate melts ($T_g \approx 450$°C) [11], using a modified Angell plot with relaxation time ($\tau$) instead of viscosity.

A linear relationship between $T_g^{\omega}$ and $\tau$ at each modulation frequency has been reported [18], and by extrapolation we can determine the glass transition temperature ($T_g^{\tau = 100}$) where $\tau$ = 100 s. By plotting $T_g^{\omega}$ reduced by $T_g^{\tau = 100}$ as a function of the $\tau$ values at each modulation frequency, we obtain an Angell plot with relaxation time instead of viscosity (Fig. 2). The definition of fragility can then be used to calculate *m*, as here done for tectosilicate calcium aluminosilicate glasses (Fig. 3). From the modulated data, *m* is found to decrease with increasing silica content, consistent with the trends observed from direct viscosity measurements and determined by standard DSC using the activation energy for structural relaxation [20]. TM-DSC thus succeeds in reproducing the composition-dependent trend in fragility, but the absolute values of *m* are systematically lower for high-*m* compositions and vice versa for low-*m* compositions. Additional studies are needed to clarify if this holds generally for all silicate glass systems and to make a correction function if necessary.

## Detecting silicate glasses with minimal relaxation

TM-DSC has been proposed to be a probe for detecting so-called "intermediate phases," featuring isostatic topology with minimal structural relaxation upon heating [21]. One of the signatures of glassy materials is their non-equilibrium state that continuously relaxes towards the supercooled liquid metastable equilibrium state. As such, a detailed understanding of the relaxation mechanism is of both scientific and industrial interest. Relaxation-free glasses can be produced by a careful design of the glass network topology, but the currently known glasses exhibiting these properties are not yet industrially relevant [22]. Therefore, there is an interest in identifying industrially useful silicate glasses with minimal structural relaxation, e.g., for high-performance display glasses. This can be achieved by using TM-DSC and identifying a minimum in the non-reversing heat flow, although the interpretation of this quantity is still under debate [23]. Fig. 4 suggests that within the fully charge-compensated calcium aluminosilicate system, the relaxation behavior can be tuned by changing the network topology, as also confirmed by supplementary molecular dynamics simulations. Furthermore, we study the magnitude of volume relaxation by annealing at a fixed viscosity (Fig. 4). Interestingly, the volume relaxation exhibits minima in the same compositional range as the local minima in non-reversing heat flow. Although more work is needed, this suggests that TM-DSC could be used as a tool to search for silicate glasses with minimal volume relaxation during heating.

## FIGURES

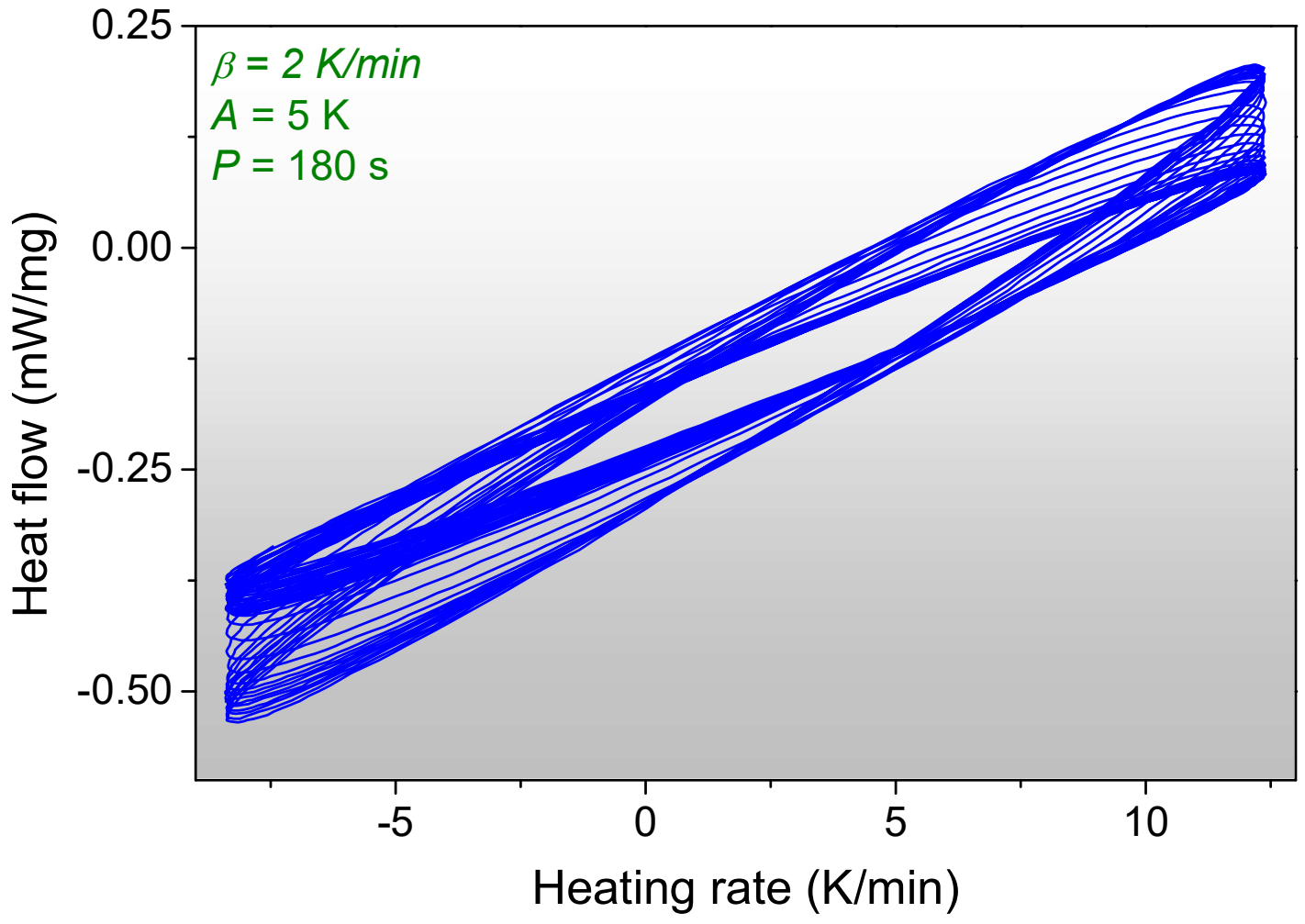

**Figure 1.** Lissajous curve for a calcium aluminosilicate glass (20CaO-25$Al_2O_3$-55$SiO_2$) with parameters: $\beta$ = 2 K/min, $A$ = 5 K, and $P$ = 150 s, showing a linear response. The data have been smoothed slightly using a Savitzky-Golay algorithm [24].

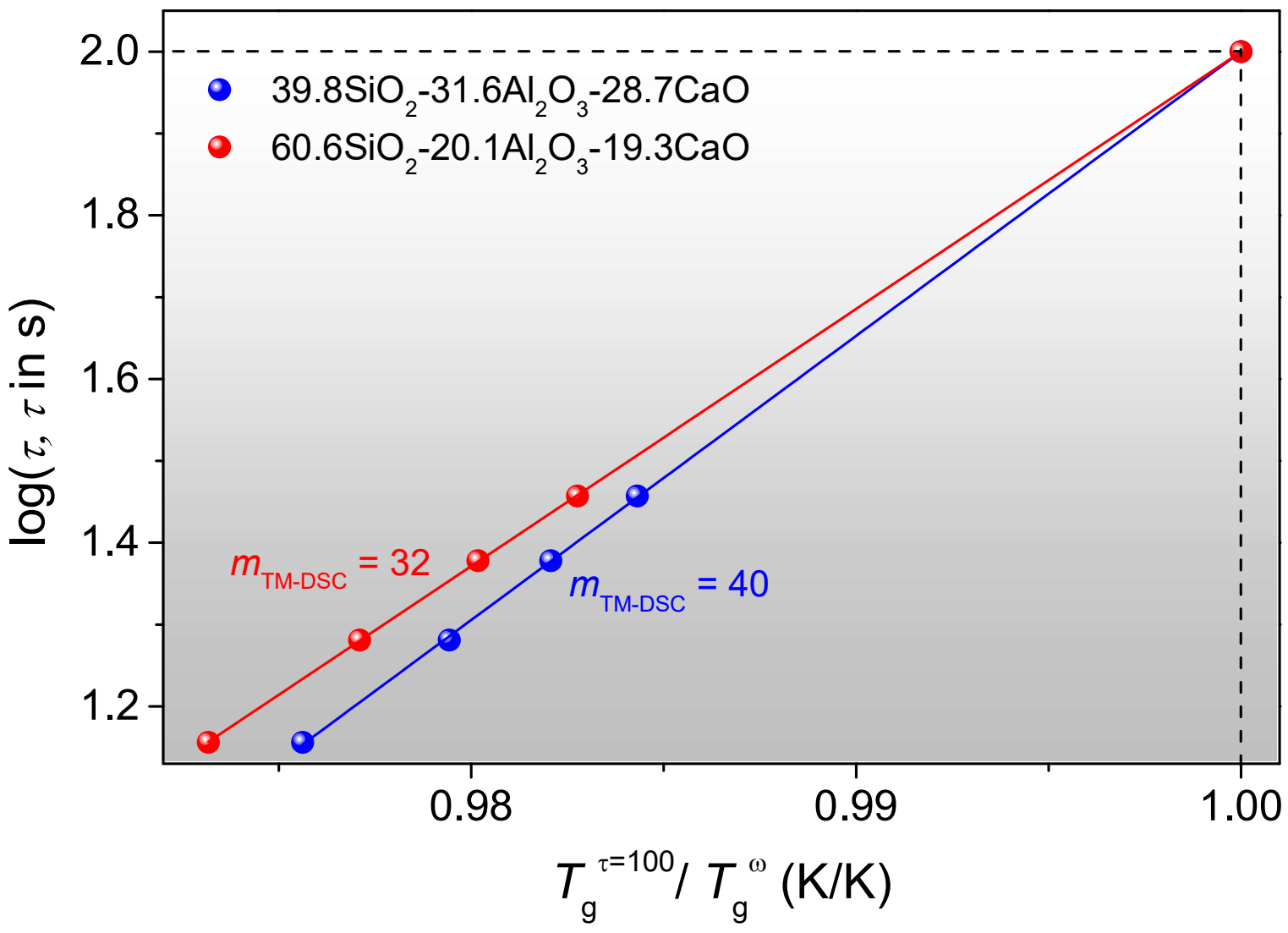

**Figure 2.** Angell plot of two calcium aluminosilicate glasses, showing the relaxation time ($\tau$) slightly above the glass transition temperature ($T_g$). The relaxation times have been determined using $\tau = 1/\omega = P/2\pi$ rad/s. The straight lines represent the best linear fit to the data.

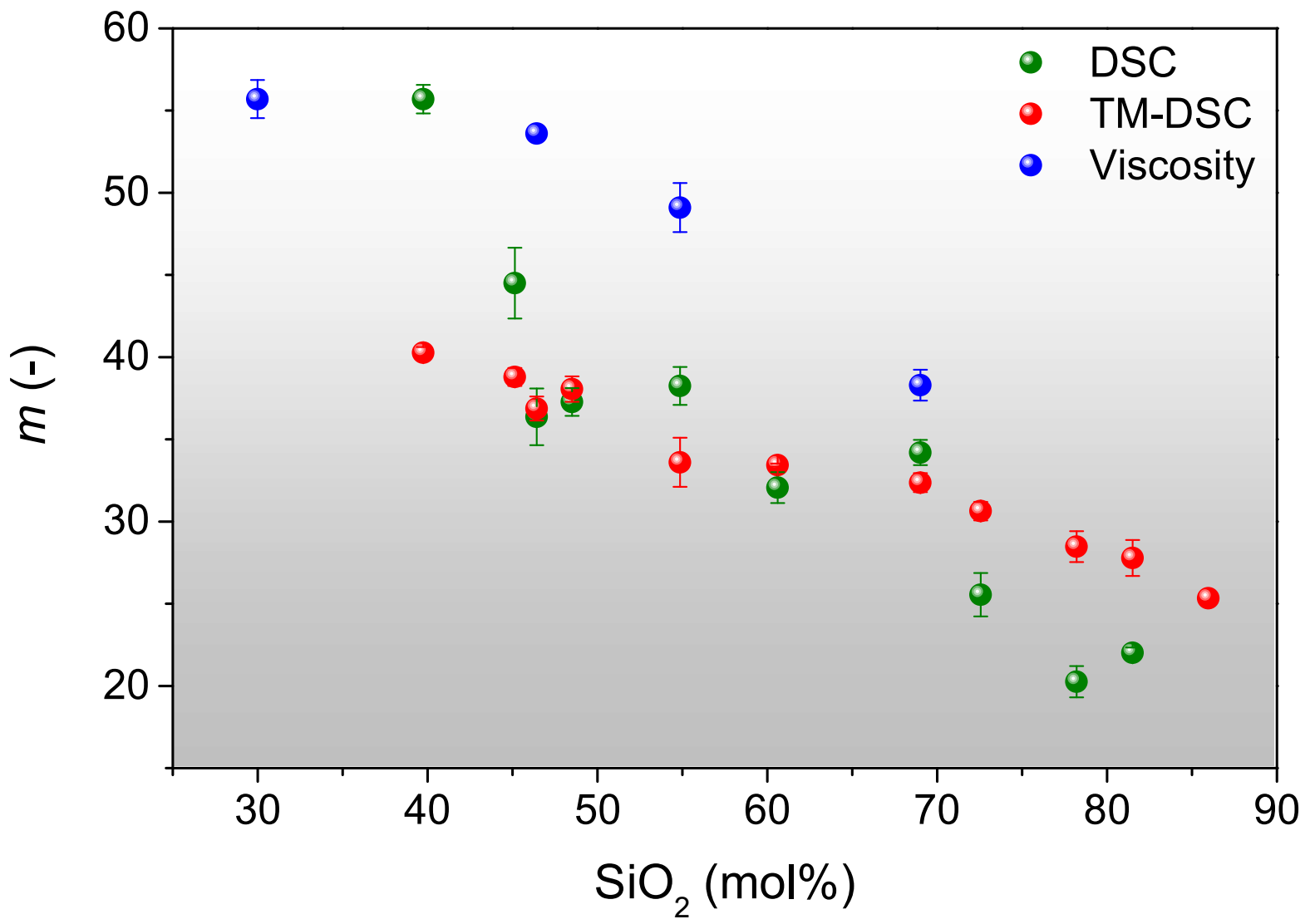

**Figure 3.** Composition dependence of the liquid fragility index *m* in fully-charge compensated calcium aluminosilicate glasses. The *m* values were determined by TM-DSC in this study, while we have determined *m* using viscometry and Moynihan's DSC procedure in a separate study [20]. The compositional trend in fragility is captured by the TM-DSC method.

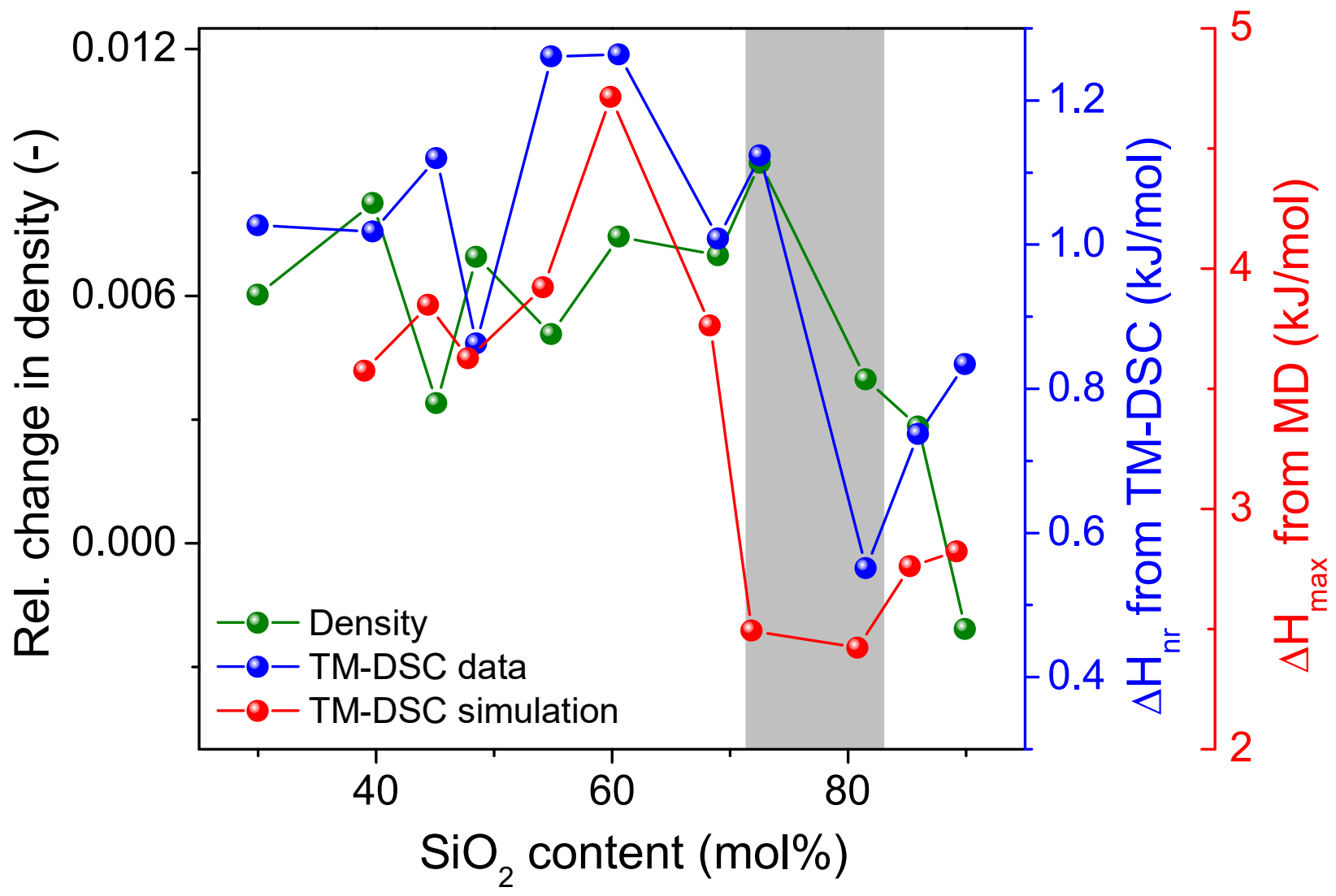

**Figure 4.** Relative change in density ($\Delta\rho/\rho_0$) in tectosilicate calcium aluminosilicates before and after sub-$T_g$ annealing at $\eta = 10^{15}$ Pa s (black). The initial density ($\rho_0$) was measured on samples annealed at $\eta = 10^{12}$ Pa s (i.e., at the standard $T_g$). The compositional dependence of non-reversing heat flow from molecular dynamics (MD) simulations and TM-DSC experiments is also shown (blue and red, respectively). The data suggest that a minimum in relaxation can be obtained by tailoring the topology of the glassy network.